\documentclass{INTERSPEECH2023}
\pdfoutput=1

\usepackage{booktabs}
\usepackage{multirow}
\usepackage{tabularx}
\usepackage{caption}
\usepackage{hyperref}
\usepackage{bm}
\usepackage{dcolumn}
\usepackage{siunitx}
\usepackage[justification=centering]{caption}
\newcolumntype{d}[1]{D{.}{.}{#1}}

\interspeechcameraready

\title{Resource-Efficient Fine-Tuning Strategies for Automatic MOS Prediction\\ in Text-to-Speech for Low-Resource Languages}

\name{Phat Do$^1$, Matt Coler$^1$, Jelske Dijkstra$^2$, Esther Klabbers$^3$}
\address{
$^1$Campus Fryslân, University of Groningen, the Netherlands\\
$^2$Fryske Akademy/Mercator Research Centre, the Netherlands\\
$^3$ReadSpeaker, the Netherlands}
\email{\{t.p.do, m.coler\}@rug.nl, jdijkstra@fryske-akademy.nl, esther.judd@readspeaker.com}

\begin{document}

\maketitle

\begin{abstract}

We train a MOS prediction model based on wav2vec 2.0 using the open-access data sets BVCC and SOMOS. Our test with neural TTS data in the low-resource language (LRL) West Frisian shows that pre-training on BVCC before fine-tuning on SOMOS leads to the best accuracy for both fine-tuned and zero-shot prediction. Further fine-tuning experiments show that using more than 30 percent of the total data does not lead to significant improvements. In addition, fine-tuning with data from a single listener shows promising system-level accuracy, supporting the viability of one-participant pilot tests. These findings can all assist the resource-conscious development of TTS for LRLs by progressing towards better zero-shot MOS prediction and informing the design of listening tests, especially in early-stage evaluation.

\end{abstract}
\noindent\textbf{Index Terms}: MOS prediction, naturalness evaluation, neural text-to-speech synthesis, low-resource languages, fine-tuning

\section{Introduction}
\label{intro}
\subsection{Mean Opinion Score (MOS) evaluation in TTS}

Neural TTS has come to dominate research in TTS. It produces better results in output speech and requires less pre-processing and feature development than concatenative and statistical parametric synthesis \cite{tanSurveyNeuralSpeech2021}. Unlike these paradigms, neural TTS generally already achieves satisfactory intelligibility, at least with sufficient training data. Thus, its evaluation is usually focused on other aspects. The most widely assessed aspect is naturalness, commonly indicated by MOS (Mean Opinion Score \cite{rec800MeanOpinion2006}) or MUSHRA (MUltiple Stimuli with Hidden Reference and Anchor \cite{2014method_mushra}) score. It should be noted that this approach is not without flaws. For example, the term ``naturalness'' itself is rather ambiguous. In fact, instead of treating TTS evaluation as a highly standardized MOS-based process as it has been, numerous studies have promoted more nuanced approaches. An example is focusing on ``contextual appropriateness'' and the product's user-centered experience \cite{wagnerSpeechSynthesisEvaluation2019a}. Nevertheless, measuring naturalness through MOS scores remains a popular method for evaluating and comparing TTS systems across studies.

Although MOS evaluation is widely used for TTS, it has its challenges. Human participants need to listen to each audio sample and rate its naturalness on a 5-point scale. As such, these listening tests require a high amount of resources, both time-wise and monetary. This is especially true if there is a large amount of samples to rate. Online listening tests do alleviate these issues, but not completely. Evaluating TTS in low-resource languages (LRLs) presents even greater challenges, as it can be difficult to find enough proficient speakers for the test and financial resources are often scarce. Combined with the typical issue of limited training data, this makes it even more challenging for TTS research in LRLs to keep pace with better-resourced languages.

\subsection{Automatic prediction of MOS}
Automatic prediction of MOS has gained attention recently as a promising solution to the challenges of traditional MOS-based evaluation. One influential work in this area is MOSNet \cite{loMOSNetDeepLearningBased2019}, which predicts MOS from magnitude spectrograms using a CNN-BLSTM architecture. Subsequent research has explored ways to improve its accuracy and generalize to different contexts. For example, \cite{williamsComparisonSpeechRepresentations2020b} experimented with eight other representations besides spectrogram frames, while MBNet \cite{lengMBNETMOSPrediction2021} and \cite{tsengUtilizingSelfSupervisedRepresentations2021} both explored explicitly learning the listener bias in the MOS data. Most recently, research effort in this topic was gathered in the VoiceMOS Challenge \cite{huangVoiceMOSChallenge20222022}, which released the BVCC data set \cite{cooperGeneralizationAbilityMOS2022}. BVCC is a collection of MOS ratings from its own large-scale listening test on samples obtained from 6 years of the Blizzard Challenge (BC) and 3 years of the Voice Conversion Challenge (VCC). BVCC was used as baseline training data in the challenge, and greatly enabled following research, e.g., \cite{manochaSpeechQualityAssessment2022, huCorrelationLossMOS2022, tianTransferMultiTaskLearning2022b}.

\subsection{MOS data set for neural TTS}
\label{intro_to_SOMOS}
BVCC consists of MOS ratings from a vast range of voice conversion (VC) and TTS systems. For the former, e.g., in VCC 2020 \cite{dasPredictionsSubjectiveRatings2020}, listeners are usually instructed to give ``naturalness'' ratings based on the converted audio quality and not the actual naturalness or pronunciation accuracy in the speech. For the latter, since the samples are from BC 2008 \cite{karaiskos2008blizzard} to BC 2016 \cite{kingBlizzardChallenge2016}, they come from a wide range of systems using vastly different designs for unit selection and/or statistical parametric synthesis. This means there is a large difference in the audio quality of the samples, and thus the listeners in BVCC likely paid more attention to this than the prosodic naturalness of the samples. This makes it different from the evaluation of neural TTS. For these reasons, \cite{maniatiSOMOSSamsungOpen2022} argues that BVCC is less suitable for training MOS prediction models aimed at neural TTS and introduces SOMOS, a new open-access data set designed for this task.

SOMOS consists of 20,100 utterances from 200 neural TTS systems based on or similar to Tacotron \cite{wangTacotronEndtoendSpeech2017}, modified to vary their prosodic manipulation ability. The authors trained MOSNet \cite{loMOSNetDeepLearningBased2019}, LDNet \cite{huangLDNetUnifiedListener2022}, and a system based on large self-supervised learning (SSL) models \cite{cooperGeneralizationAbilityMOS2022} on SOMOS, and showed promising results on its test set. However, there were no results on external test sets, which would provide a clearer comparison between BVCC and SOMOS. We address this directly here.

Furthermore, to the best of our knowledge, there is currently no work on MOS prediction that specifically focuses on TTS for LRLs. Due to the typical lack of training data for the acoustic model, TTS for LRLs may present its own problematic patterns in the speech prosody (or naturalness in general). These patterns are uncharacteristic in TTS for better-resourced languages and are thus not picked up by general MOS prediction models. Additionally, there is a need for LRL-specific fine-tuning strategies that focus on efficiency, given the constraints on available resources. Therefore, we investigate strategies for fine-tuning MOS prediction models for LRLs, particularly regarding the necessary amount of data and fine-tuning with data from a single listener.

\subsection{Contributions}
We make the following contributions:
\begin{itemize}
    \item[1)]{We use two large open-access MOS data sets (BVCC and SOMOS) to train a prevalent SSL-based MOS prediction architecture and compare their performance with neural TTS samples in the LRL West Frisian without any fine-tuning.}

	\item[2)]{We fine-tune the models above with different amounts (10\% to 40\%) of total data to explore a ``sweet spot'' in the amount of fine-tuning data, which may give insights that can inform resource-efficient listening test design for TTS in LRLs.}

	\item[3]{Lastly, we experiment with fine-tuning using MOS data from only one listener, to explore the feasibility of using MOS data from a single listener to fine-tune TTS models in LRLs.}

\end{itemize}

  \begin{figure*}[!h]
  \includegraphics[width=\linewidth]{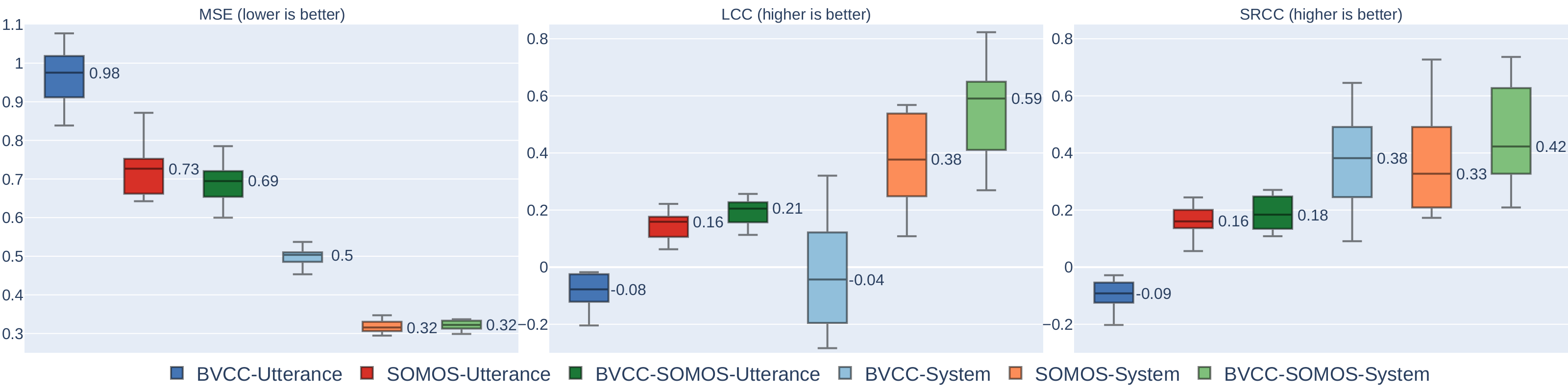}
  \vspace{-5mm}
  \caption{Zero-shot prediction performance in MSE, LCC, and SRCC, at utterance and system levels (from 10 data splits)}
  \vspace{-2mm}
  \label{zeroshot_figure}
  \end{figure*}

\section{Experiment details}
\subsection{MOS prediction model}
\label{MOS_prediction_system}
Large self-supervised learning (SSL) models for speech have been proved to have strong performance when fine-tuned for various speech-related tasks, and \cite{cooperGeneralizationAbilityMOS2022} followed this approach for MOS prediction. They took several pre-trained SSL models from the FAIRSEQ project \cite{ottFairseqFastExtensible2019}, added a linear layer to the models' output embeddings, and fine-tuned them using L1 loss calculated from MOS ratings. One of their best performing models was fine-tuned from \textit{wav2vec 2.0 Base} (95M parameters) \cite{baevskiWav2vecFrameworkSelfSupervised2020}. This was later used as one of the baselines in the VoiceMOS Challenge \cite{huangVoiceMOSChallenge20222022} and turned out to be among the best performing systems in the out-of-domain (OOD) track. OOD prediction matches our interest since it also has to use limited data. This model also had the best accuracy in the recent experiments with SOMOS in \cite{maniatiSOMOSSamsungOpen2022}. Therefore, we chose this model's architecture for the current study. We followed its authors' recommendations in audio pre-processing (e.g., resampling to \SI{16}{\kilo\hertz}, normalizing amplitude with \textit{sv56} \cite{internationaltelecommunicationunionRecommendation191Software}). We also kept the authors' hyper-parameters for stochastic gradient descent: learning rate $\num{1e-4}$, momentum $0.9$, and early stopping after 20 epochs of no decrease in validation loss.

\subsection{MOS data set}
\label{MOS_frisian_dataset}
We used MOS data from our previous work \cite{doTexttoSpeechUnderResourcedLanguages2022}, which proposed a new phone mapping method in transfer learning for TTS in LRLs and experimented with 30 minutes of data in the LRL West Frisian (hereafter Frisian). We used FastSpeech 2 \cite{renFastSpeechFastHighQuality2023} for the acoustic model and HiFi-GAN V1 \cite{kongHiFiGANGenerativeAdversarial2020} for the vocoder, which makes it suitable for our focus on neural TTS for LRLs. We experimented on fine-tuning different source language acoustic models, with and without phone mapping. There were in total 220 synthetic samples from 11 systems: 10 TTS systems and 1 hidden resynthesized anchor, following the MUSHRA guideline \cite{2014method_mushra}. Frisian speakers listened to each sample and rated its pronunciation accuracy and naturalness using the 0-100 MUSHRA scale. We converted the naturalness ratings linearly to the 5-scale of MOS and used them for the current study. In total, there were 2,024 individual MOS ratings from 46 listeners. The synthetic samples, listening scores, and relevant participant information are available online\footnote{\url{phat-do.github.io/sigul22}}.

This data set was split into training/validation/test sets of 40\%/10\%/50\%. We ensured a relatively even distribution of TTS systems among them. Due to the rather small data size, we wanted to avoid chance findings and to reduce the effect of too influential ratings. Therefore, we did the random split 10 times, each resulting in a different combination of the three sets. For the following steps, we repeated the work for all 10 splits.

\subsection{Model training and prediction}
Using the architecture in Section~\ref{MOS_prediction_system}, we compared three scenarios: \textit{a)} training on BVCC, \textit{b)} training on SOMOS, and \textit{c)} pre-training on BVCC then training further on SOMOS. Hereafter we refer to these as \textit{BVCC}, \textit{SOMOS}, and \textit{BVCC-SOMOS}, respectively. We used the author-provided checkpoint for \textit{BVCC} \cite{cooperGeneralizationAbilityMOS2022} and trained the other two using the ``clean'' subset of SOMOS. We checked the performance of \textit{SOMOS} (on its own test set) against the original paper to verify that its training was in order. These three resulting models were then tested on the Frisian test sets in Section~\ref{MOS_frisian_dataset} without fine-tuning (zero-shot). The results are shown in Figure~\ref{zeroshot_figure} and discussed in Section~\ref{zeroshot_results}.

We then fine-tuned these models with different amounts of data. We divided the 40\% training set into four subsets, each containing a different proportion of the total data (10\%, 20\%, 30\%, and 40\%), and fine-tuned the models using these subsets. At each interval, we used the checkpoint for prediction. This was repeated for each of the 10 splits. The results are shown in Figure~\ref{finetuned_figure} and discussed in Section~\ref{finetuned_results}. All training was done with a batch size of 4 on a single NVIDIA V100 32GB GPU, and all fine-tuning was done with batch size 1 on an NVIDIA GTX 1080 (8GB) GPU. The training of \textit{SOMOS} and \textit{BVCC-SOMOS} each took roughly 7.5 hours.

The fine-tuning above followed the convention in previous studies, but this is not always viable, especially for TTS in LRLs. This is because although it required fewer rated test samples, these samples still came from the same number of listeners as the whole data set. In other words, researchers would still need to find the same number of listeners, only that each listener would need to rate fewer samples. Although such fine-tuning is obviously still advantageous, another more likely scenario for LRLs is that researchers could find a few enthusiastic listeners that would not mind rating many samples, rather than the other way around. Besides, the work by \cite{pineRequirementsMotivationsLowResource2022a} shows that, for LRLs with very small communities, the evaluation of a few community-engaged and respected speakers of the language can be representative of that of the whole community.

With this in mind, we wanted to see how the models would perform when fine-tuned on ratings from only one listener. We picked out 21 listeners that completed rating their assigned set of 44 samples (i.e., 20\% of the total number of samples) and had self-rated Frisian proficiency and usage frequency at the highest level. We then fine-tuned using these 21 mini sets and compared with the corresponding all-listener model from above (also fine-tuned with 20\% data, 10\% for training and 10\% for validation). Their performance on the same test set was compared, with results in Figure~\ref{1listener_figure} and discussion in Section~\ref{1listener_results}.

For all scenarios, we analyzed three accuracy metrics, all calculated between ground-truth and predicted ratings: MSE (Mean Squared Error), LCC (Linear Correlation Coefficient, i.e., Pearson's \textit{r}), and SRCC (Spearman Rank Correlation Coefficient). Both LCC and SRCC were used as we were interested in both the linear correlation and the monotonic relationship (correlation in ranking order). All metrics were calculated at the utterance level and system level (averaging the predicted ratings per TTS system). Where relevant, all calculations were repeated for the 10 random splits as mentioned in Section~\ref{MOS_frisian_dataset}.

\section{Results \& discussion}
\subsection{Zero-shot prediction}
\label{zeroshot_results}

As seen in Figure~\ref{zeroshot_figure}, similar to the findings of \cite{cooperGeneralizationAbilityMOS2022} and \cite{maniatiSOMOSSamsungOpen2022}, system-level prediction generally performed better than the utterance-level. This is very likely because the latter is more affected by the varied listener bias. Also, rather similar to \cite{maniatiSOMOSSamsungOpen2022}, here \textit{SOMOS} had better accuracy than \textit{BVCC}, confirming that the SOMOS data set's focus on neural TTS made it more suitable as training data for predicting this type of TTS.

More interestingly, \textit{BVCC-SOMOS} outperformed \textit{BVCC} and \textit{SOMOS} in almost all cases. To test the statistical significance of this result, we used Wilcoxon ranked tests between \textit{BVCC-SOMOS} and its runner-up, with different alternative hypotheses depending on the metric. In the case where \textit{BVCC-SOMOS} tied with \textit{SOMOS} in system MSE, we conducted a separate Wilcoxon test to confirm that their medians were not different from each other. For this case, we compared \textit{BVCC-SOMOS} with \textit{BVCC} instead. The results in Table~\ref{zeroshot_pvalue} confirm that \textit{BVCC-SOMOS} had the best accuracy among the three. 

There are two likely explanations for this result. First, it is possible that the limited TTS training data led to artifacts or patterns that are somewhat characteristic of BVCC (mentioned in Section~\ref{intro_to_SOMOS}). Second, pre-training on BVCC may have provided an advantage due to its varied audio source types.

\begin{table}[!h]
  \vspace{-1mm}
  \footnotesize
  \caption{Wilcoxon tests for BVCC-SOMOS\\ outperforming others at zero-shot prediction}
  \vspace{-1mm}
  \centering
  \begin{tabular}{@{}cccrc@{}}
  \toprule
  \textbf{Metric} & \textbf{Level} & \textbf{Alt. hypo.} & \textbf{\textit{p}-value} & \textbf{Sig. code} \\ \midrule
  \multirow{2}{*}{MSE}                & Utterance                          & $<$ SOMOS                                     & $.006$ & **                                \\
                                      & System                             & $<$ BVCC $~~$                                      & $<.001$ & ***                     \\ \midrule
  \multirow{2}{*}{LCC}                & Utterance                          & $>$ SOMOS                                  & $.018$ & *                 \\
                                      & System                             & $>$ SOMOS                                  & $.001$ & **                                \\ \midrule
  \multirow{2}{*}{SRCC}               & Utterance                          & $>$ SOMOS                                  & $.065$ & .                                \\
                                      & System                             & $>$ BVCC $~~$                                 & $.050$ & *                               \\
  \bottomrule
  \end{tabular}
  \label{zeroshot_pvalue}
  \vspace{-2mm}
  \end{table}

\begin{figure*}[!ht]
  \includegraphics[width=\linewidth]{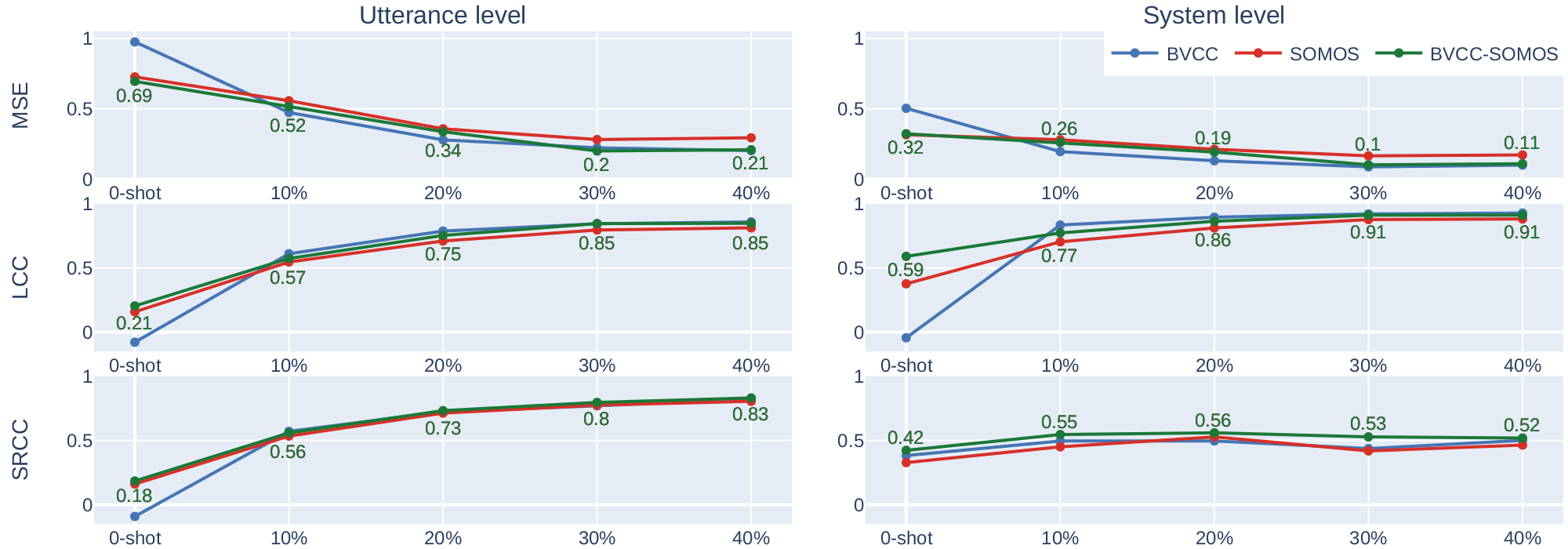}
  \vspace{-6mm}
  \caption{Fine-tuned performance from different amounts of fine-tuning data (BVCC-SOMOS median values are annotated)}
  \label{finetuned_figure}
  \vspace{-2mm}
  \end{figure*}

Additionally, \textit{BVCC-SOMOS}' moderate-high LCC at the system level ($M=0.59, \max=0.82$) hints that it can be used for zero-shot prediction, at least to a certain extent. For example, it can be useful in comparing different TTS systems (different techniques or model variants) being developed before organizing a real listening test. This is particularly valuable for LRLs. The lower SRCC could be explained by the small number (11) of TTS systems, meaning any error in the predicted ranking was heavily penalized. For reference, the data from the Blizzard Challenge 2019 \cite{wu2019blizzard} (BC2019, the OOD data in \cite{huangVoiceMOSChallenge20222022}) has 26 systems.

\subsection{Fine-tuned prediction}
\label{finetuned_results}

Figure~\ref{finetuned_figure} shows that \textit{BVCC-SOMOS} and \textit{BVCC} generally outperformed \textit{SOMOS}, and Wilcoxon ranked tests between the former two showed significant differences in only 3 out of 24 scenarios. Therefore, for simplicity, the rest of the discussion will focus on \textit{BVCC-SOMOS}, as it performed the best in zero-shot (Section~\ref{zeroshot_results}) and arguably co-best in fine-tuned prediction.

We conducted Wilcoxon ranked tests of the accuracy metrics between each interval of increase in fine-tuning data to find a ``sweet spot'' for the amount of fine-tuning data against performance. For instance, we compared the accuracy metrics from zero-shot to 10\%, and from 10\% to 20\%. The results showed that all accuracy measures stopped improving after the 30\%-step, i.e., adding training data beyond this produced no statistically significant improvement. Different experiment settings would likely result in a different ``pivot point'', but this finding can be of value for similar experiments. The only exception was system SRCC, where the intervals led to no improvement. This may be again related to the small amount of TTS systems, but could be investigated in future work.

For validation in another test language (Chinese), we compared \textit{BVCC-SOMOS} with \textit{BVCC} using BC2019. We used the test set (40\%) and training set (10\%) as provided. Table~\ref{compare_BC2019} shows that \textit{BVCC-SOMOS} again mostly outperformed \textit{BVCC}\footnote{Where relevant, we used results from \cite{huangVoiceMOSChallenge20222022}, which only ran each test once, so we could not use a statistical test for more robust results.}. 

\subsection{One-listener prediction}
\label{1listener_results}

\begin{figure*}[!ht]
  \includegraphics[width=\linewidth]{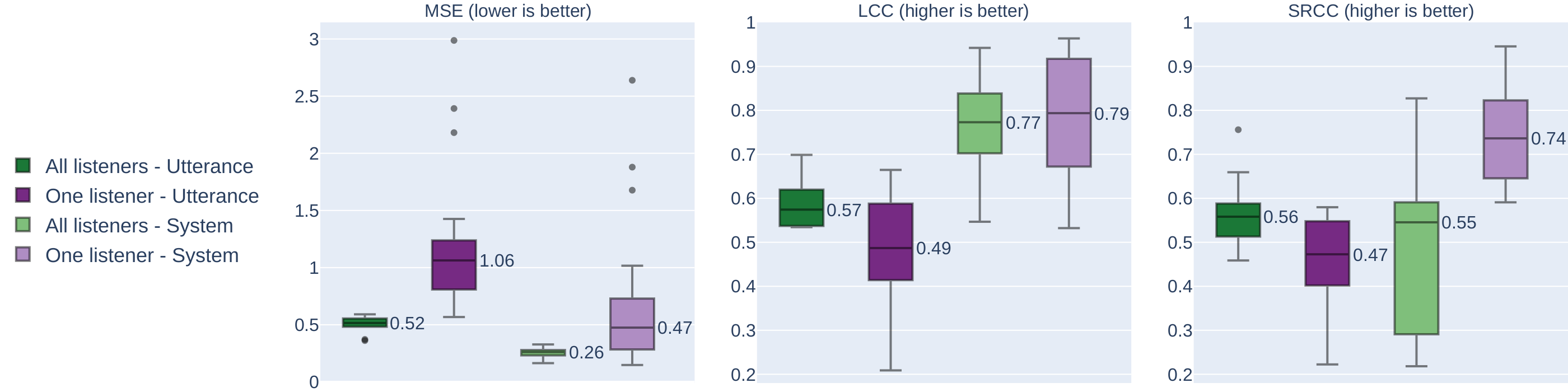}
  \vspace{-6mm}
  \caption{Prediction performance of fine-tuned BVCC-SOMOS: all-listener vs. one-listener data (at 20\% of total data)}
  \vspace{-2mm}
  \label{1listener_figure}
  \end{figure*}

\begin{table}[!h]
  \centering
  \footnotesize
  \caption{Comparison with BVCC using BC2019 data}
  \vspace{-1mm}
  \addtolength{\tabcolsep}{-0.3em}
  \begin{tabular}{@{}cccccc@{}}
  \toprule
  \multirow{2}{*}{\textbf{Metric}} & \multirow{2}{*}{\textbf{Level}} & \multicolumn{2}{c}{\textbf{Zero-shot}} & \multicolumn{2}{c}{\textbf{Fine-tuned}} \\ \cmidrule(l){3-6} 
                                    & & BVCC-SOMOS  & BVCC   & BVCC-SOMOS   & BVCC   \\ \midrule
  \multirow{2}{*}{MSE} & Utt                          & \textbf{1.801}       & $3.924$           & \textbf{0.245}        & $0.260$           \\
    & Sys                          & \textbf{1.803}       & $3.967$           & \textbf{0.065}        & $0.099$           \\ \midrule
  \multirow{2}{*}{LCC} & Utt                          & \textbf{0.468}       & $0.394$           & \textbf{0.895}        & $0.888$           \\
    & Sys                          & \textbf{0.560}       & $0.419$           & \textbf{0.975}        & $0.971$           \\ \midrule
  \multirow{2}{*}{SRCC} & Utt                         & $0.394$                & \textbf{0.464}  & \textbf{0.859}        & $0.849$           \\
    & Sys                         & $0.438$                & \textbf{0.549}  & $0.973$                 & \textbf{0.975}  \\ \bottomrule
  \end{tabular}
  \label{compare_BC2019}
  \vspace{-6mm}
  \end{table}

As expected, fine-tuning with ratings from one listener was much more challenging, as shown in Figure~\ref{1listener_figure}. The one-listener models did much worse and occasionally resulted in very high outliers in MSE, likely due to the highly varying listener bias. However, system-level LCC and SRCC were interestingly higher than in the all-listener models. This could be because the one-listener data had ratings for all TTS systems from the same listener. This helped the prediction model learn the relative ranking order of the TTS systems even better than the all-listener model (which had more ``noise''). In other words, the predicted ratings were more different from the ground-truth (shown through the high MSEs), but the relative order of the TTS systems was closer to the ground-truth. This shows the viability of using a pilot one-participant listening test in early comparison of multiple TTS systems being developed. This is particularly valuable for TTS in LRLs.

\subsection{Assumptions \& limitations}
Most of our experiments were in Frisian, but the validation in Chinese (Table~\ref{compare_BC2019}) also gave largely similar results. Since Frisian and Chinese are very different, similar results can be expected for other languages. However, the assumption that similar results would be obtained on other languages remains a limitation because of the limited scope and the absence of MOS data in other languages. Further research is required to address this limitation. In addition, we assumed that linearly converting the MUSHRA scores to MOS was appropriate for the analysis.

\section{Conclusions}

Our study compared two large open-access MOS datasets (BVCC and SOMOS) using a prevalent SSL-based MOS prediction architecture, and evaluated their performance in predicting MOS for LRLs. We found that pre-training on BVCC and training further on SOMOS led to the best results, both in zero-shot and fine-tuned prediction. Obtaining and fine-tuning with around 30\% of the total MOS data was identified as a useful strategy for pilot listening tests in developing TTS for LRLs, or in more general low-resource settings (the ``sweet spot''). Additionally, we found that using ratings from only one listener could be a promising early-stage way to compare different TTS models in development.

Our experiments in Frisian and Chinese suggest that similar results could be achieved in other languages, although this remains an assumption that requires further validation. Moving forward, it will be important to extend these findings to a wider range of data sets and explore new evaluation metrics, such as contextual appropriateness, to continue advancing our understanding of how to best develop TTS for LRLs.

In summary, our study highlights effective strategies for predicting MOS and identifies potential areas for future research in TTS for LRLs. By applying these findings in practice, we can work toward improving the development of TTS systems for speakers of low-resource languages.

\section{Acknowledgements}
We thank the Center for Information Technology of the University of Groningen for their support and for providing access to the Hábrók high performance computing cluster.

\newpage

\bibliographystyle{IEEEtran}

\bibliography{mybib}

\begin{thebibliography}{10}
\providecommand{\url}[1]{#1}
\csname url@samestyle\endcsname
\providecommand{\newblock}{\relax}
\providecommand{\bibinfo}[2]{#2}
\providecommand{\BIBentrySTDinterwordspacing}{\spaceskip=0pt\relax}
\providecommand{\BIBentryALTinterwordstretchfactor}{4}
\providecommand{\BIBentryALTinterwordspacing}{\spaceskip=\fontdimen2\font plus
\BIBentryALTinterwordstretchfactor\fontdimen3\font minus
  \fontdimen4\font\relax}
\providecommand{\BIBforeignlanguage}[2]{{%
\expandafter\ifx\csname l@#1\endcsname\relax
\typeout{** WARNING: IEEEtran.bst: No hyphenation pattern has been}%
\typeout{** loaded for the language `#1'. Using the pattern for}%
\typeout{** the default language instead.}%
\else
\language=\csname l@#1\endcsname
\fi
#2}}
\providecommand{\BIBdecl}{\relax}
\BIBdecl

\bibitem{tanSurveyNeuralSpeech2021}
X.~Tan, T.~Qin, F.~Soong, and T.-Y. Liu, ``A {{Survey}} on {{Neural Speech
  Synthesis}},'' \emph{arXiv:2106.15561 [cs, eess]}, Jun. 2021.

\bibitem{rec800MeanOpinion2006}
I.~Rec, ``P. 800.1, {{Mean}} {Opinion} {Score} ({{MOS}}) {Terminology},''
  \emph{International Telecommunication Union, Geneva}, 2006.

\bibitem{2014method_mushra}
``{Method} for the {subjective assessment} of {intermediate quality level} of
  {audio systems},'' \emph{International Telecommunication Union
  Radiocommunication Assembly}, 2014.

\bibitem{wagnerSpeechSynthesisEvaluation2019a}
P.~Wagner, J.~Beskow, S.~Betz, J.~Edlund, J.~Gustafson, G.~Eje~Henter,
  S.~Le~Maguer, Z.~Malisz, {\'E}.~Sz{\'e}kely, C.~T{\aa}nnander, and
  J.~Vo{\ss}e, ``Speech {{Synthesis Evaluation}} \textemdash{}
  {{State-of-the-Art Assessment}} and {{Suggestion}} for a {{Novel Research
  Program}},'' in \emph{10th {{ISCA Workshop}} on {{Speech Synthesis}} ({{SSW}}
  10)}.\hskip 1em plus 0.5em minus 0.4em\relax {ISCA}, Sep. 2019, pp. 105--110.

\bibitem{loMOSNetDeepLearningBased2019}
C.-C. Lo, S.-W. Fu, W.-C. Huang, X.~Wang, J.~Yamagishi, Y.~Tsao, and H.-M.
  Wang, ``{{MOSNet}}: {{Deep Learning-Based Objective Assessment}} for {{Voice
  Conversion}},'' in \emph{Interspeech 2019}.\hskip 1em plus 0.5em minus
  0.4em\relax {ISCA}, Sep. 2019, pp. 1541--1545.

\bibitem{williamsComparisonSpeechRepresentations2020b}
J.~Williams, J.~Rownicka, P.~Oplustil, and S.~King, ``Comparison of {{Speech
  Representations}} for {{Automatic Quality Estimation}} in {{Multi-Speaker
  Text-to-Speech Synthesis}},'' in \emph{The {{Speaker}} and {{Language
  Recognition Workshop}} ({{Odyssey}} 2020)}.\hskip 1em plus 0.5em minus
  0.4em\relax {ISCA}, Nov. 2020, pp. 222--229.

\bibitem{lengMBNETMOSPrediction2021}
Y.~Leng, X.~Tan, S.~Zhao, F.~Soong, X.-Y. Li, and T.~Qin, ``{{MBNET}}: {{MOS
  Prediction}} for {{Synthesized Speech}} with {{Mean-Bias Network}},'' in
  \emph{{{ICASSP}} 2021 - 2021 {{IEEE International Conference}} on
  {{Acoustics}}, {{Speech}} and {{Signal Processing}} ({{ICASSP}})}, Jun. 2021,
  pp. 391--395.

\bibitem{tsengUtilizingSelfSupervisedRepresentations2021}
W.-C. Tseng, C.-y. Huang, W.-T. Kao, Y.~Y. Lin, and H.-y. Lee, ``Utilizing
  {{Self-Supervised Representations}} for {{MOS Prediction}},'' in
  \emph{Interspeech 2021}.\hskip 1em plus 0.5em minus 0.4em\relax {ISCA}, Aug.
  2021, pp. 2781--2785.

\bibitem{huangVoiceMOSChallenge20222022}
W.~C. Huang, E.~Cooper, Y.~Tsao, H.-M. Wang, T.~Toda, and J.~Yamagishi, ``The
  {{VoiceMOS Challenge}} 2022,'' in \emph{Interspeech 2022}.\hskip 1em plus
  0.5em minus 0.4em\relax {ISCA}, Sep. 2022, pp. 4536--4540.

\bibitem{cooperGeneralizationAbilityMOS2022}
E.~Cooper, W.-C. Huang, T.~Toda, and J.~Yamagishi, ``Generalization {{Ability}}
  of {{MOS Prediction Networks}},'' in \emph{{{ICASSP}} 2022 - 2022 {{IEEE
  International Conference}} on {{Acoustics}}, {{Speech}} and {{Signal
  Processing}} ({{ICASSP}})}, May 2022, pp. 8442--8446.

\bibitem{manochaSpeechQualityAssessment2022}
P.~Manocha and A.~Kumar, ``Speech {{Quality Assessment}} through {{MOS}} using
  {{Non-Matching References}},'' in \emph{Interspeech 2022}.\hskip 1em plus
  0.5em minus 0.4em\relax {ISCA}, Sep. 2022, pp. 654--658.

\bibitem{huCorrelationLossMOS2022}
B.~Hu and Q.~Li, ``Correlation {{Loss}} for {{MOS Prediction}} of {{Synthetic
  Speech}},'' in \emph{2022 {{Asia-Pacific Signal}} and {{Information
  Processing Association Annual Summit}} and {{Conference}} ({{APSIPA ASC}})},
  Nov. 2022, pp. 1910--1914.

\bibitem{tianTransferMultiTaskLearning2022b}
X.~Tian, K.~Fu, S.~Gao, Y.~Gu, K.~Wang, W.~Li, and Z.~Ma, ``A {{Transfer}} and
  {{Multi-Task Learning}} based {{Approach}} for {{MOS Prediction}},'' in
  \emph{Interspeech 2022}.\hskip 1em plus 0.5em minus 0.4em\relax {ISCA}, Sep.
  2022, pp. 5438--5442.

\bibitem{dasPredictionsSubjectiveRatings2020}
R.~K. Das, T.~Kinnunen, W.-C. Huang, Z.-H. Ling, J.~Yamagishi, Z.~Yi, X.~Tian,
  and T.~Toda, ``Predictions of {{Subjective Ratings}} and {{Spoofing
  Assessments}} of {{Voice Conversion Challenge}} 2020
  {{Submissions}}\vphantom\{\}\vphantom\{\},'' in \emph{Proc. {{Joint
  Workshop}} for the {{Blizzard Challenge}} and {{Voice Conversion Challenge}}
  2020}, 2020, pp. 99--120.

\bibitem{karaiskos2008blizzard}
V.~Karaiskos, S.~King, R.~Clark, and C.~Mayo, ``{The Blizzard Challenge
  2008},'' in \emph{Proc. Blizzard Challenge Workshop}, 2008.

\bibitem{kingBlizzardChallenge2016}
S.~King and V.~Karaiskos, ``The {{Blizzard Challenge}} 2016,'' in \emph{Proc.
  Blizzard Challenge Workshop}, 2016.

\bibitem{maniatiSOMOSSamsungOpen2022}
G.~Maniati, A.~Vioni, N.~Ellinas, K.~Nikitaras, K.~Klapsas, J.~S. Sung, G.~Jho,
  A.~Chalamandaris, and P.~Tsiakoulis, ``{{SOMOS}}: {{The Samsung Open MOS
  Dataset}} for the {{Evaluation}} of {{Neural Text-to-Speech Synthesis}},'' in
  \emph{Interspeech 2022}.\hskip 1em plus 0.5em minus 0.4em\relax {ISCA}, Sep.
  2022, pp. 2388--2392.

\bibitem{wangTacotronEndtoendSpeech2017}
Y.~Wang, R.~{Skerry-Ryan}, D.~Stanton, Y.~Wu, R.~J. Weiss, N.~Jaitly, Z.~Yang,
  Y.~Xiao, Z.~Chen, S.~Bengio, Q.~Le, Y.~Agiomyrgiannakis, R.~Clark, and R.~A.
  Saurous, ``Tacotron: {{Towards End-to-End Speech Synthesis}},'' in
  \emph{Interspeech 2017}.\hskip 1em plus 0.5em minus 0.4em\relax {ISCA}, Aug.
  2017, pp. 4006--4010.

\bibitem{huangLDNetUnifiedListener2022}
W.-C. Huang, E.~Cooper, J.~Yamagishi, and T.~Toda, ``{{LDNet}}: {{Unified
  Listener Dependent Modeling}} in {{MOS Prediction}} for {{Synthetic
  Speech}},'' in \emph{{{ICASSP}} 2022 - 2022 {{IEEE International Conference}}
  on {{Acoustics}}, {{Speech}} and {{Signal Processing}} ({{ICASSP}})}, May
  2022, pp. 896--900.

\bibitem{ottFairseqFastExtensible2019}
M.~Ott, S.~Edunov, A.~Baevski, A.~Fan, S.~Gross, N.~Ng, D.~Grangier, and
  M.~Auli, ``Fairseq: {{A Fast}}, {{Extensible Toolkit}} for {{Sequence
  Modeling}},'' in \emph{Proceedings of the 2019 {{Conference}} of the {{North
  American Chapter}} of the {{Association}} for {{Computational Linguistics}}
  ({{Demonstrations}})}.\hskip 1em plus 0.5em minus 0.4em\relax {Minneapolis,
  Minnesota}: {Association for Computational Linguistics}, Jun. 2019, pp.
  48--53.

\bibitem{baevskiWav2vecFrameworkSelfSupervised2020}
A.~Baevski, Y.~Zhou, A.~Mohamed, and M.~Auli, ``Wav2vec 2.0: {{A Framework}}
  for {{Self-Supervised Learning}} of {{Speech Representations}},'' in
  \emph{Advances in {{Neural Information Processing Systems}}}, vol.~33.\hskip
  1em plus 0.5em minus 0.4em\relax {Curran Associates, Inc.}, 2020, pp.
  12\,449--12\,460.

\bibitem{internationaltelecommunicationunionRecommendation191Software}
I.~T. Union, ``Recommendation {{G}}.191: {{Software}} tools for speech and
  audio coding standardization,''
  https://www.itu.int/rec/T-REC-G.191-202207-I/en.

\bibitem{doTexttoSpeechUnderResourcedLanguages2022}
P.~Do, M.~Coler, J.~Dijkstra, and E.~Klabbers, ``Text-to-{{Speech}} for
  {{Under-Resourced Languages}}: {{Phoneme Mapping}} and {{Source Language
  Selection}} in {{Transfer Learning}},'' in \emph{Proceedings of the 1st
  {{Annual Meeting}} of the {{ELRA}}/{{ISCA Special Interest Group}} on
  {{Under-Resourced Languages}}}.\hskip 1em plus 0.5em minus 0.4em\relax
  {Marseille, France}: {European Language Resources Association}, Jun. 2022,
  pp. 16--22.

\bibitem{renFastSpeechFastHighQuality2023}
Y.~Ren, C.~Hu, X.~Tan, T.~Qin, S.~Zhao, Z.~Zhao, and T.-Y. Liu,
  ``{{FastSpeech}} 2: {{Fast}} and {{High-Quality End-to-End Text}} to
  {{Speech}},'' in \emph{International {{Conference}} on {{Learning
  Representations}}}, Feb. 2023.

\bibitem{kongHiFiGANGenerativeAdversarial2020}
J.~Kong, J.~Kim, and J.~Bae, ``{{HiFi-GAN}}: {{Generative Adversarial
  Networks}} for {{Efficient}} and {{High Fidelity Speech Synthesis}},'' in
  \emph{Advances in {{Neural Information Processing Systems}}}, vol.~33.\hskip
  1em plus 0.5em minus 0.4em\relax {Curran Associates, Inc.}, 2020, pp.
  17\,022--17\,033.

\bibitem{pineRequirementsMotivationsLowResource2022a}
A.~Pine, D.~Wells, N.~Brinklow, P.~Littell, and K.~Richmond, ``Requirements and
  {{Motivations}} of {{Low-Resource Speech Synthesis}} for {{Language
  Revitalization}},'' in \emph{Proceedings of the 60th {{Annual Meeting}} of
  the {{Association}} for {{Computational Linguistics}} ({{Volume}} 1: {{Long
  Papers}})}.\hskip 1em plus 0.5em minus 0.4em\relax {Dublin, Ireland}:
  {Association for Computational Linguistics}, May 2022, pp. 7346--7359.

\bibitem{wu2019blizzard}
Z.~Wu, Z.~Xie, and S.~King, ``{The Blizzard Challenge 2019},'' in \emph{Proc.
  Blizzard Challenge Workshop}, vol. 2019, 2019.

\end{thebibliography}

\end{document}